%% file: rac1p.tex
\documentclass{llncs}

\usepackage{amssymb}
\usepackage{graphicx}
\usepackage{subfigure}
\usepackage{xspace}
\usepackage{wrapfig}

\newtheorem{observation}{Observation}

\title{1-Bend RAC Drawings of 1-Planar Graphs
\thanks{Research supported in part by the MIUR
    project AMANDA ``Algorithmics for MAssive and Networked DAta'',
    prot. 2012C4E3KT\_001.}}

\author{Walter Didimo\inst{1} \and
Giuseppe Liotta\inst{1} \and
Saeed Mehrabi\inst{2} \and
\\Fabrizio Montecchiani\inst{1}
}

\institute{
Dipartimento di Ingegneria, Universit{\`a} degli Studi di Perugia, Italy\\
\email{ \{walter.didimo,giuseppe.liotta,fabrizio.montecchiani\}@unipg.it}
\and
David R. Cheriton School of Computer Science, University of Waterloo, Canada\\
\email{smehrabi@uwaterloo.ca}}

\begin{document}

\maketitle

\begin{abstract}
A graph is \emph{$1$-planar} if it has a drawing where each edge is crossed at most once. A drawing is \emph{RAC (Right Angle Crossing)} if the  edges cross only at right angles. The relationships between $1$-planar graphs and RAC drawings have been partially studied in the literature. It is known that there are both $1$-planar graphs that are not straight-line RAC drawable and  graphs that have a straight-line RAC drawing but that are not $1$-planar~\cite{el-rac1p-DAM13}. Also, straight-line RAC drawings always exist for \emph{IC-planar graphs}~\cite{DBLP:journals/tcs/BrandenburgDEKL16}, a subclass of $1$-planar graphs. 
%An IC-planar graph is a graph that admits a $1$-planar drawing with independent crossings (i.e., no two crossed edges share an endpoint). 
One of the main questions still open is whether every $1$-planar graph has a RAC drawing with at most one bend per edge. We positively answer this question.
\end{abstract}

\section{Introduction}
An emerging research line in Graph Drawing studies families of non-planar graphs that can be drawn so that crossing edges verify some desired properties. This topic is informally recognized as ``beyond planarity''. Different types of properties give rise to different families of beyond planar graphs. Among them, particular attention has been devoted to \emph{$1$-planar graphs}~(see, e.g.,~\cite{DBLP:journals/dam/Ackerman14,DBLP:conf/gd/AlamBK13,SoCG,JGAA-330,DBLP:journals/tcs/BrandenburgDEKL16,DBLP:journals/ipl/Didimo13,DBLP:journals/algorithmica/GrigorievB07,help-ft1pg-COCOON12,DBLP:journals/jgt/KorzhikM13,pt-gdfce-C97,r-esadk-AMS65}) and to \emph{RAC (Right Angle Crossing) graphs} (see, e.g.,~\cite{DBLP:journals/cj/ArgyriouBS13,DBLP:journals/jgaa/BekosDKW16,DBLP:journals/algorithmica/GiacomoDEL14,DBLP:journals/cj/DiGiacomoDGLR15,DBLP:journals/comgeo/GiacomoDLM13,del-dgrac-2011,dl-cargd-12,dlr-tdfda-10,DBLP:journals/jgaa/DidimoLR11,nehh-lcacl-2010,DBLP:conf/vl/HuangEHL10}). A graph is $1$-planar if it has a drawing where each edge is crossed at most once, while it is RAC if it has a polyline drawing where the edges cross only at right angles. From an application point of view, the study of these two families is motivated by several cognitive experiments, suggesting that the readability of a layout is negatively correlated to the number of crossings~\cite{DBLP:journals/iwc/Purchase00,DBLP:journals/ese/PurchaseCA02,DBLP:journals/ivs/WarePCM02} and that user task performances are not affected too much if edges cross at large angles~\cite{DBLP:conf/apvis/Huang07,DBLP:journals/vlc/HuangEH14,DBLP:conf/apvis/HuangHE08}. Also, users often prefer straight-line drawings or layouts whose edges have few bends~\cite{DBLP:conf/gd/Purchase97}, and several algorithms optimize this aesthetic criterion~\cite{dett-gd-99}.  Note that, every graph admits a polyline RAC drawing with at most three bends per edge~\cite{del-dgrac-2011}. 

For the reasons above, it is interesting to study what graphs can be drawn with at most one crossing per edge, right angle crossings, and few bends per edge at the same time. We recall that $n$-vertex $1$-planar graphs have at most $4n-8$ edges~\cite{pt-gdfce-C97} and that straight-line $1$-planar drawings have at most $4n-9$ edges~\cite{DBLP:journals/ipl/Didimo13}. Also, straight-line RAC graphs have at most $4n-10$ edges~\cite{del-dgrac-2011}, while RAC drawings with at most one bend per edge or two bends per edge, have at most $6.5n-13$ and $74.2n$ edges, respectively~\cite{DBLP:journals/comgeo/ArikushiFKMT12}. These results immediately imply that there are $1$-planar graphs not admitting $1$-planar drawings with straight-line edges and $1$-planar graphs not admitting straight-line drawings with right angle crossings. Also, there exist straight-line RAC drawable graphs that are not $1$-planar~\cite{el-rac1p-DAM13}. 
In this scenario, one of the main questions still open is whether every $1$-plane graph admits a RAC drawing with at most one bend per edge. This paper positively answers this question, by proving the following result. 

\begin{theorem}\label{th:main}
Let $G$ be an $n$-vertex 1-planar graph. Then $G$ admits a 1-planar RAC drawing $\Gamma$ with at most one bend per edge. Also, if a 1-planar embedding of $G$ is given as part of the input, $\Gamma$ can be computed in $O(n)$ time.
\end{theorem}

We remark that a characterization of the $1$-planar graphs that can be drawn with straight-line edges was given by Thomassen in 1988~\cite{t-rdg-JGT88}. The characterization is described in terms of the existence of a $1$-planar embedding that does not contain two primitive forbidden configurations. This result immediately implies that every $1$-planar graph admits a $1$-planar drawing with at most one bend per edge (which is not necessarily RAC); it is sufficient to subdivide each crossing edge of any given $1$-planar embedding with a dummy vertex, so to remove any possible forbidden configuration. Dummy vertices will correspond to bends in the final drawing. Moreover, Alam et al.~\cite{DBLP:conf/gd/AlamBK13}, proved that every $3$-connected $1$-plane graph can be drawn with straight-line edges, except for at most one edge that may require one bend. We also remark that straight-line RAC drawings always exist for \emph{IC-planar graphs}~\cite{DBLP:journals/tcs/BrandenburgDEKL16}, a subclass of $1$-planar graphs. 
%An IC-planar graph is a graph that admits a $1$-planar drawing with independent crossings (i.e., no two crossed edges share an end-vertex). 

Some proofs and technicalities can be found in the appendix.

\section{Preliminaries}\label{se:preliminaries}

We assume familiarity with basic terminology of graph drawing~\cite{dett-gd-99}. In the following we only consider \emph{simple} drawings of graphs, i.e., drawings where two edges have at most one point in common (which is either a common endpoint or a common interior point where the two edges properly cross each other). A \emph{$k$-bend drawing} of a graph is a drawing  where each edge is represented as a polyline with at most $k > 0$ bends.
A graph $G$ is \emph{planar} if it admits a planar (i.e., crossing-free) drawing. Such a drawing subdivides the plane into topologically connected regions, called \emph{faces}. The infinite region is the \emph{outer face}. The number of vertices encountered in the closed walk along the boundary of a face $f$ is the \emph{degree} of $f$. If $G$ is not $2$-connected a vertex may be encountered more than once, thus contributing with more than one unit to the degree of $f$. A \emph{planar embedding} of $G$ is an equivalence class of planar drawings of $G$ having the same set of faces. A \emph{plane graph} is a planar graph with a given planar embedding.  

The concept of planar embedding can be extended to non-planar drawings. Given a non-planar drawing $\Gamma$, interpret every crossing as a vertex. The resulting planarized drawing has a planar embedding. An \emph{embedding} of a (non-planar) graph $G$ is an equivalence class of  drawings whose planarized versions have the same planar embedding. A \emph{1-plane} graph is a 1-planar graph with a given \emph{1-planar embedding}, i.e., an embedding where each edge is crossed at most once. 
Each face of a 1-planar embedding is composed of both vertices and/or crossings, and its degree is the number of vertices or crossings  encountered in the closed walk along its boundary. 
%
%\begin{figure}[t]
%\centering
%\includegraphics[scale=1]{figures/triangulated-example}
%\caption{\label{fi:triangulated-example}A triangulated 1-plane graph.}
%\end{figure}
%
A \emph{kite} $K$ is a 1-plane graph isomorphic to $K_4$ with an embedding such that all the vertices are on the boundary of the outer face, the four edges on the boundary are crossing-free, and the remaining two edges cross each other. Given a 1-plane graph $G$ and a kite $K=\{a,b,c,d\}$, such that $K \subseteq G$, we say that $K$  is \emph{empty} if it does not contain any vertex of $G$ inside the 4-cycle $\{a,b,c,d\}$ (it contains only the crossing point). 
A pair of crossing edges of $G$ \emph{forms an empty kite} if their four end-vertices induce an empty kite. A 1-plane graph $G$, possibly containing parallel edges, is \emph{triangulated} if each face is a triangle, formed by either three vertices or by one crossing and two vertices. Clearly, a triangulated 1-plane graph is 2-connected.  The next observation follows from the definition of a triangulated 1-plane graph (see Fig.~\ref{fi:triangulated-example} in the appendix for an example).

\begin{observation}\label{ob:emptykite}
Let $G$ be a triangulated 1-plane graph. Every pair of crossing edges of $G$ forms an empty kite, except for at most one pair of crossing edges if their crossing point is on the outer face of $G$. 
\end{observation}

%For example, Fig.~\ref{fi:triangulated-example} shows a triangulated 1-plane graph with two pairs of crossing edges. One of these two pairs  forms an empty kite, while the other one does not since its crossing is part of the outer face.

\section{1-bend RAC Drawings of 1-planar Graphs}\label{se:1bend}
\begin{figure}[t]
\centering
\subfigure[$G$]{\includegraphics[scale=0.8,page=1]{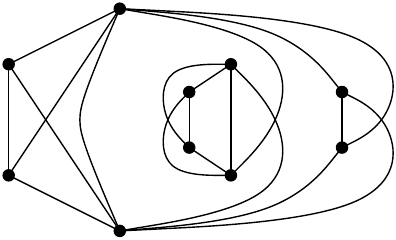}\label{fi:g}}\hfill
\subfigure[$G_2$]{\includegraphics[scale=0.8,page=2]{figures/algorithm}\label{fi:g2}}\hfill
\subfigure[$G^+$]{\includegraphics[scale=0.8,page=3]{figures/algorithm}\label{fi:gplus}}
\caption{Illustration for the augmentation step.}
\end{figure}

To prove Theorem~\ref{th:main} we give an algorithm that takes as input a simple 1-plane graph $G$ with $n$ vertices (see, e.g., Fig.~\ref{fi:g}), and computes a 1-bend 1-planar RAC drawing $\Gamma$ of $G$ in $O(n)$ time. 
We assume that $G$ is connected, as otherwise we can draw independently each connected component. 
The high-level idea is as follows. First augment $G$ and modify its embedding to get a triangulated 1-plane graph, possibly containing parallel edges. Then, execute a suitable decomposition of the graph and apply a recursive technique that computes a 1-bend 1-planar RAC drawing.

%\subsection{Augmentation}\label{sse:augmentation} 

\paragraph{Augmentation.} The first step of the algorithm transforms $G$ into a triangulated 1-plane graph $G^+$ by adding edges and vertices. The 1-planar embedding of $G^+$ may be different from that of $G$ for the common part.  
%
%\begin{figure}
%\centering
%\subfigure[]{\includegraphics[scale=1,page=1]{figures/augmentation}\label{fi:k4}}\hfil
%\subfigure[]{\includegraphics[scale=1,page=2]{figures/augmentation}\label{fi:emptyreg}}\hfil
%\caption{Illustration for the augmentation step: (a) Crossing augmentation; (b) Edge $e_2$ is removed.}
%\end{figure}
%
Let $(a,c)$ and $(b,d)$ be two edges of $G$ that cross in a point $p$.  Let $\{a,b,c,d\}$ be the circular order of the vertices around $p$. For each such pair of crossing edges, we add an edge $(a,b)$, and draw\footnote{For ease of description, here we are interpreting an embedding as a drawing.} it such that it follows the curves $(a,p)$ and $(p,b)$. Similarly, we draw the three edges $(b,c)$, $(c,d)$ and $(d,a)$ (see also Fig.~\ref{fi:k4} in the appendix).  
This operation ensures that each pair of crossing edges forms an empty kite. Also, this operation does not introduce edge crossings but it may create parallel edges. We denote by $G_1$ the resulting (multi)graph. For each pair of parallel edges $e$ and $e'$ of $G_1$, such that $e \in G$ and $e' \in G_1$, we remove $e$ from $G_1$. This immediately implies that no parallel edge is crossed in $G_1$. We then remove one edge for each pair of parallel edges $e_1$ and $e_2$ such that the curve $e_1 \cup e_2$ does not contain any vertex in its interior (see Fig.~\ref{fi:emptyreg} in the appendix). 
We let $G_2$ be the resulting graph, which can be easily computed in $O(n)$ time, since $G$ has $O(n)$ crossings (see, e.g.,~\cite{Suzuki2010}). Figure~\ref{fi:g2} shows the graph $G_2$ obtained from the graph $G$ of Fig.~\ref{fi:g}. We remark that a similar operation has been used by Alam {et al.}~\cite{DBLP:conf/gd/AlamBK13} in order to compute a straight-line drawable 1-planar embedding of a 3-connected 1-planar graph. However, only 3-connected graphs are considered by Alam et al., and in this case the augmented graph does not contain parallel edges~\cite{DBLP:conf/gd/AlamBK13}. We do not have any restriction on the connectivity of $G$,  which poses additional issues in the construction  and in the drawing of a suitable 1-planar embedding. We transform $G_2$ into a triangulated 1-plane graph. Note that a face of degree two consists of two parallel edges, thus only the outer face of $G_2$ can have degree two. In this case, each of the two parallel edges is part of an empty kite. Thus,  we remove one of these two edges to make the degree of the outer face equal to three (it will be formed by two vertices and one crossing).  Let $f$ be an inner face of $G_2$ that is not a triangle. Such a face contains no crossings on its boundary, since each crossing is shared by exactly four triangular faces by the empty kite property. We add an \emph{extra vertex} $v_f$ inside $f$ and connect it to all vertices (with multiplicity) on the boundary of $f$. Figure~\ref{fi:gplus} shows the graph $G^+$ obtained from the graph $G_2$ in Fig.~\ref{fi:g2}, extra vertices are drawn as squares. Since $G_2$ has $O(n)$ faces, $G^+$ has $O(n)$ vertices and edges, and it is computed in $O(n)$ time.  
The next lemma follows from the above discussion.

\begin{lemma}\label{le:g2}
Graph $G^+$ is a triangulated 1-plane (multi)graph. 
\end{lemma}

%\subsection{Decomposition}\label{sse:decomposition}

\paragraph{Decomposition.} We define a decomposition of $G^+$ inspired by $SPQR$-trees~\cite{DBLP:journals/siamcomp/BattistaT96}, but simpler and more direct for our purposes. The next lemma can be proved.
%The next lemma is proved in the appendix.

\begin{lemma}\label{le:seppair}
Let $G$ be a triangulated 1-plane (multi)graph and let $\{u,v\}$ be a separation pair of $G$. 
There exist two parallel edges $e, e'$ incident to $u$ and $v$ such that $\{u,v\}$ is not a separation pair for the graph obtained by removing from $G$ all vertices inside the cycle $\{e, u, e', v\}$. 
\end{lemma}

\begin{figure}[t]
\centering
\subfigure[$G^+$]{\includegraphics[scale=0.8,page=4]{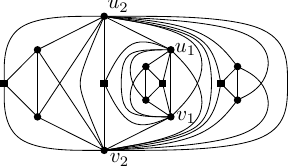}\label{fi:gplus-2}}\hfill
\subfigure[]{\includegraphics[scale=0.8,page=5]{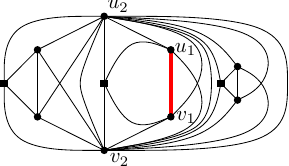}\label{fi:decomp-1}}\hfill
\subfigure[$G^*$]{\includegraphics[scale=0.8,page=6]{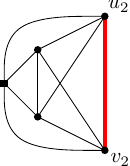}\label{fi:decomp-2}}
\caption{Illustration for the decomposition step. Thick edges are thicker (and red). }
\end{figure}

By Lemma~\ref{le:seppair}, for each separation pair $\{u,v\}$ of $G^+$, there exist $k>1$ parallel edges $\{e_1, \dots, e_k\}$  between $u$ and $v$, such that the cycle $\{u, e_1, v, e_k\}$ encloses all other copies in its interior. We call the \emph{inner graph}  of $(u,v)$ the subgraph  $G_{uv}$ of $G^+$ whose outer face is $\{u, e_1, v, e_k\}$, and an \emph{inner component}  of $(u,v)$ each subgraph $C^i_{uv}$ of $G_{uv}$ whose outer face is $\{u, e_i, v, e_{i+1}\}$, for $i=1,\dots,k-1$. Let $G_{uv}$ be an inner graph of $G^+$ that does not contain any inner graph as a subgraph. Replace $G_{uv}$ with an edge between $u$ and $v$, called \emph{thick edge}; the resulting graph is still a triangulated 1-plane graph.  Iterate this procedure until there are no more inner graphs to be replaced. This is done in $O(n)$ time and results in a simple triangulated 1-plane graph $G^*$, which is 3-connected by Lemma~\ref{le:seppair}. Figure~\ref{fi:decomp-2} shows the graph $G^*$ obtained from the graph $G^+$ in Fig.~\ref{fi:gplus-2}, through the intermediate step in Fig.~\ref{fi:decomp-1}. 
The next lemma follows.

\begin{lemma}\label{le:inner}
Graph $G^*$ is a simple 3-connected triangulated 1-plane graph.
\end{lemma}

%\subsection{Drawing}\label{sse:drawing}

\begin{figure}[t]
\centering
\subfigure[$H^*$]{\includegraphics[scale=0.8,page=7]{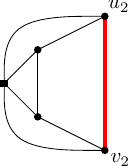}\label{fi:hstar}}\hfill
\subfigure[$\gamma^*$]{\includegraphics[scale=0.8,page=8]{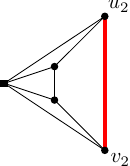}\label{fi:gammastar}}\hfill
\subfigure[$\Gamma^*$]{\includegraphics[scale=0.8,page=9]{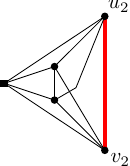}\label{fi:gammastar-2}}\hfill
\subfigure[]{\includegraphics[scale=0.8,page=10]{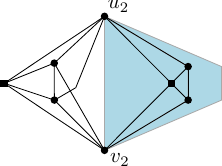}\label{fi:hk}}\hfill
\subfigure[]{\includegraphics[scale=0.8,page=11]{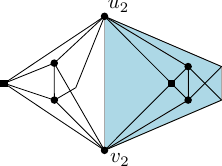}\label{fi:hk-2}}\hfill
\subfigure[]{\includegraphics[scale=0.8,page=12]{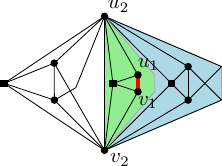}\label{fi:hi}}\hfill
\subfigure[]{\includegraphics[scale=0.8,page=13]{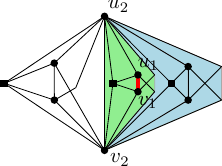}\label{fi:hi-2}}\hfill
\subfigure[]{\includegraphics[scale=0.8,page=14]{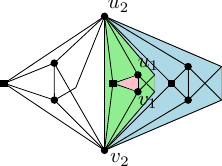}\label{fi:hi-3}}\hfill
\subfigure[]{\includegraphics[scale=0.8,page=15]{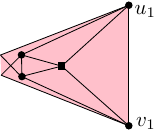}\label{fi:hi-4}}
\caption{Illustration for the drawing step. }
\end{figure}

\paragraph{Drawing.} The overview of the drawing algorithm is as follows. Start with a 1-bend 1-planar RAC drawing of $G^*$, and then recursively replace thick edges with a 1-bend 1-planar RAC drawing of the corresponding inner graphs. Deleting the edges and vertices added by the augmentation step we get a 1-bend 1-planar RAC drawing of $G$. 
To compute a 1-bend 1-planar RAC drawing of $G^*$, first remove from $G^*$ all pairs of crossing edges and denote by $H^*$ the resulting plane graph (see Fig.~\ref{fi:hstar}). Note that thick edges are never crossed by construction, and all faces of $H^*$ have either degree 3 or degree 4. %We also prove the following.
We can prove the following.
%The following lemma is proved in the appendix.

\begin{lemma}\label{le:3conn}
Graph $H^*$ is 3-connected.
\end{lemma}

Compute a planar straight-line drawing $\gamma^*$ of $H^*$ where all faces are strictly convex and the outer face is a prescribed polygon $P$; this can be done by applying the linear-time algorithm by Chiba et al.~\cite{chiba1984linear} (see Fig.~\ref{fi:gammastar}). If the outer face of $H^*$ has degree four, we let $P$ be a trapezoid, else $P$ is a triangle. Since all faces are either triangles or quadrangles, we can avoid three collinear vertices by slight perturbations (which cannot cause a face to become non convex). To reinsert the crossing edges, we distinguish between the inner faces and the outer face of $H^*$. Two crossing edges can be easily reinserted in an inner face, just drawing one of the two with no bend and the other with one bend, such that they cross at right angles (see, e.g.,~\cite{DBLP:journals/jgaa/AngeliniCDFBKS11} and Fig.~\ref{fi:gammastar-2}). To reinsert two crossing edges $e_1, e_2$ in the outer face of $H^*$ so that they form a right angle, we can draw $e_1$ and $e_2$ with one bend each (see also Fig.~\ref{fi:outerface} in the appendix). 
Namely, $P$ is a trapezoid by construction. Assume that the minor base $m$ and the greater base $M$ of $P$ are aligned with the horizontal axis. The first segment of $e_1$ is such that its rightmost endpoint $p_1$ coincides with the rightmost endpoint of $m$, and its leftmost endpoint $q_1$ is $b$ units above the leftmost endpoint of $m$, where $b$ is equal to the length of $m$. The second segment of $e_1$ has $q_1$ as rightmost endpoint, and its leftmost endpoint $r_1$ coincides with the leftmost endpoint of $M$. Edge $e_2$ is drawn symmetrically.
%Similarly, the first segment of $e_2$ is such that its leftmost endpoint $p_2$ coincides with the leftmost endpoint of $m$, and its rightmost endpoint $q_2$ is $b$ units above the rightmost endpoint of $m$. The second segment of $e_2$ has $q_2$ as leftmost endpoint, and its rightmost endpoint $r_2$ coincides with the rightmost endpoint of $M$.  
%This ensures that the two edges cross at right angles. 

%\begin{figure}
%\centering
%\includegraphics[scale=1]{figures/outerface}
%\caption{\label{fi:outerface}Reinserting the pair of crossing edges on the outer face of $H^*$. }
%\end{figure} 
% 

Consider now a thick edge $(u,v)$ of $G^*$ and its inner graph $G_{uv}$.  Recall that $G_{uv}$ consists of $k-1 \geq 1$ inner components $C_1,\dots,C_{k-1}$. Each $C_i$ ($i=1,\dots,k-1$) has two parallel edges $e_i$, $e_{i+1}$ as outer face. Also, analogously to $G^*$, $C^-_i = C_i \setminus \{e_{i+1}\}$ is a simple 3-connected triangulated 1-plane graph (it is a subgraph of $G^+$ and all its inner graphs have been replaced by thick edges). Remove all crossing edges of $C^-_{k-1}$ and let $H^-_{k-1}$ be the resulting 3-connected plane graph. Compute a planar straight-line drawing $\gamma_{k-1}$ of $H^-_{k-1}$ such that all faces are strictly convex polygons and the outer face is a prescribed polygon $P$. If the outer face of $H^-_{k-1}$ has degree three, $P$ is a triangle whose side with corners $u$ and $v$ has length equal to the length of the thick edge $(u,v)$ in $\Gamma^*$, and its height is small enough so that the thick edge $(u,v)$ can be replaced with $P$ without introducing crossings. If the outer face of $H^-_{k-1}$ has degree four, $P$ is a trapezoid such that its greater base has $u$ and $v$ as corners and the same length as the thick edge $(u,v)$ in $\Gamma^*$. The height of $P$ is such that the thick edge $(u,v)$ can be replaced with $P$ without introducing crossings.  Also, the minor base of $P$ is sufficiently short so that the pair of crossing edges on the outer face of $H^-_{k-1}$ can be reinserted without introducing crossings in $\Gamma^*$, as described for $H^*$ (see Fig.~\ref{fi:hk}). By the same argument used for $H^*$, all pairs of crossing edges can be reinserted so as to form right angle crossings and have at most one bend each (see Fig.~\ref{fi:hk-2}).   
If $k-1>1$, we iterate this procedure and compute a drawing $\Gamma^-_i$ for each $C^-_i$, for $i=k-2,\dots,1$. The polygon representing the outer face of each $\Gamma_i$ can be suitably chosen so to fit inside the face containing edge $e_{i+1}$ of drawing of $\Gamma_{i+1}$. The union of all such drawings is a 1-bend 1-planar RAC drawing $\Gamma_{uv}$ of $G_{uv}$ (see Figs.~\ref{fi:hi} and~\ref{fi:hi-2}), with the exception of some parallel edges. Namely, the parallel edges $e_1,\dots,e_k$ are represented by overlapping segments between $u$ and $v$, and for our needs all of them but one can be removed from the drawing. 

Repeat this procedure for each thick edge of $G^*$, and recursively apply the same technique for each inner graph of $G^*$; see Figs.~\ref{fi:hi-3} and~\ref{fi:hi-4} for a complete illustration. The resulting drawing $\Gamma$ is a 1-bend 1-planar RAC drawing of $G^+$ (except for some parallel edges). Removing dummy vertices and edges, we get the desired drawing of $G$. In terms of time complexity, each planar straight-line drawing with (strictly) convex faces is computed in linear time in the size of the input graph~\cite{chiba1984linear}, and in linear time we can reinsert the crossing edges. Thus the whole procedure takes $O(n)$ time.
This concludes the proof of Theorem~\ref{th:main}.

\section{Conclusions and Open Problems}
We proved that every 1-planar graph admits a 1-planar RAC drawing with at most one bend per edge. 
The proof is constructive and based on a drawing algorithm, which may produce 1-bend 1-planar RAC drawings with exponential area: Is this area requirement necessary for some 1-planar graphs?  Also, our algorithm may change the embedding of the input graph: Are there 1-planar embeddings that are not realizable as 1-bend RAC drawings?
Characterizing straight-line 1-planar RAC drawable graphs is also an interesting problem.

%Our algorithm may produce drawings with exponential area in the size of the input graph: Is this area requirement sometimes necessary for 1-planar RAC drawings with at most one bend per edge? 
%Also, our algorithm may change the embedding of the input graph: Are there 1-planar embeddings that cannot be realized as RAC drawings with at most one bend per edge?
%Finally, the  problem of finding a complete characterization of what graphs admit a bendless 1-planar RAC drawing is still open.

{\small \bibliography{rac1p}}
\bibliographystyle{splncs03}

\clearpage

\input{appendix}

\end{document}

%% file: appendix.tex
\appendix

\section*{Appendix}

\section{Additional Material for Section~\ref{se:preliminaries}}

\begin{figure}
\centering
\includegraphics[scale=0.9]{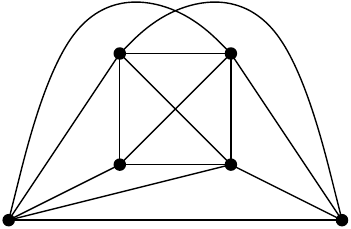}
\caption{\label{fi:triangulated-example}A triangulated 1-plane graph with two pairs of crossing edges. One of these two pairs  forms an empty kite, while the other one does not since its crossing is part of the outer face.}
\end{figure}

\section{Additional Material for Section~\ref{se:1bend}}

\begin{figure}
\centering
\subfigure[]{\includegraphics[scale=0.9,page=1]{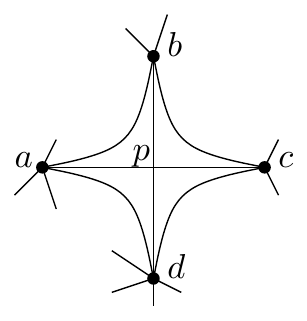}\label{fi:k4}}\hfil
\subfigure[]{\includegraphics[scale=0.9,page=2]{figures/augmentation}\label{fi:emptyreg}}\hfil
\caption{Illustration for the augmentation step: (a) Crossing augmentation; (b) Edge $e_2$ is removed.}
\end{figure}

\begin{figure}[h]
\centering
\includegraphics[scale=0.9,page=1]{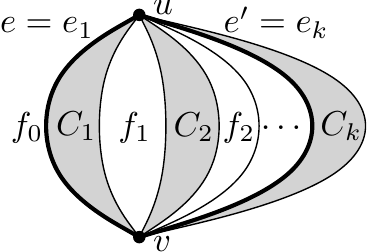}
\caption{\label{fi:triangulated}Illustration for the proof of Lemma~\ref{le:seppair}.}
\end{figure}
\paragraph*{Proof of Lemma~\ref{le:seppair}.} 
Refer to Fig.~\ref{fi:triangulated}. 
Let $C_1$, $\dots$, $C_k$, be the $k \geq 2$ connected components obtained by removing $u$ and $v$ from $G$. Observe first that there is no pair of crossing edges $e_i$, $e_j$, such that $e_i \in C_i$ and $e_j \in C_j$, with $1 \leq i \neq j \leq k$, as otherwise, by Observation~\ref{ob:emptykite}, $C_i$ and $C_j$ would not be distinct components. It follows that $C_1$, $\dots$, $C_k$ can be ordered (and possibly relabeled) such that for every pair of indices $i$ and $j$ with $1 \leq i < j \leq k$, $C_i$ is encountered before $C_j$ when walking clockwise around $v$, starting from any edge of $C_1$ incident to $v$ (see also Fig.~\ref{fi:triangulated}). Denote by $f_i$, for $i=1,\dots,k$, the face of $G$ between $C_i$ and $C_{i+1}$ (indices taken modulo $k$). Since all faces are triangles, each face $f_i$ contains, besides $u$ and $v$, either a vertex distinct from $u$ and $v$, or a crossing. It follows that there are $k$ parallel edges between $u$ and $v$, one for each face $f_i$. Denote by $e=e_1$ the parallel edge between $u$ and $v$ in $f_1$, and by $e'=e_k$ the parallel edge between $u$ and $v$ in $f_k$ (they are bold in the example of Fig.~\ref{fi:triangulated}). Then, the graph obtained by removing all vertices inside the cycle $\{e, u, e', v\}$ corresponds either to  $C_1$, or to $C_{k}$, and, in both cases, the statement holds.
\qed

\begin{figure}
\centering
\includegraphics[scale=0.9]{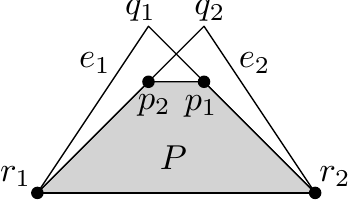}
\caption{\label{fi:outerface}Reinserting the pair of crossing edges on the outer face of $H^*$. }
\end{figure} 

\paragraph*{Proof of Lemma~\ref{le:3conn}.} Recall that $H^*$ is simple. Clearly, $H^*$ is connected, as every pair of crossing edges in $G^*$ forms an empty kite and the removal of two crossing edges cannot disconnect the graph. 

Suppose, for a contradiction, that $H^*$ contains a cut vertex $c$. Then there is a face $f$ of $H^*$ such that $c$ is encountered at least twice in a closed walk $\mathcal C$ along the boundary of $f$. Consider two consecutive occurrences, denoted by $c_1$ and $c_2$, of $c$ in $\mathcal C$. If in $\mathcal C$ no further vertex is encountered between $c_1$ and $c_2$, then $c$ has a self-loop, which is not possible. If only one vertex $v$ is encountered, then either $v$ has degree one in $H^*$, or there are two parallel edges between $c$ and $v$, and both cases are not possible. Hence, between any two consecutive 
occurrences of $c$ in $\mathcal C$ there must be at least two distinct vertices, and thus the degree of $f$ is at least six, a contradiction with the fact that all faces of $H^*$ have degree either three or four.

It remains to show that $H^*$ contains no separation pair. 
Suppose, for a contradiction, that $H^*$ contains a separation pair $\{u,v\}$. 
Then there are at least two faces of $H^*$, denoted by $f_1$ and $f_2$, such that $f_1$ and $f_2$ share no edge and contain both $u$ and $v$. Either both $f_1$ and $f_2$ have degree four, or one has degree three and the other one has degree four. Since every face of degree four of $H^*$ corresponds to a kite in $G^*$, in both cases $G^*$ contains at least two parallel edges between $u$ and $v$, a contradiction with the fact that $G^*$ is simple.
\qed